# Strain Assessment in Graphene Through the Raman 2D' Mode

*Elena del Corro, Ladislav Kavan, Martin Kalbac, Otakar Frank*[*]

J.Heyrovsky of Institute of Physical Chemistry of the AS CR, v.v.i., Dolejskova 2155/3, 182 23 Prague 8, Czech Republic.

ABSTRACT

Accurate and simple local strain assessment in graphene is one of the crucial tasks in device characterization. Raman spectroscopy is often used for that purpose through monitoring of the G and 2D modes. However, the shifts of those two bands might be biased, especially under uniaxial strain, by the effects of charge-transfer doping. Therefore, it is extremely desirable to use another Raman band, less affected by doping, but with a defined and measurable behavior under strain. The Raman 2D' mode is in this sense the ideal feature for the evaluation of strain levels in stretched graphene monolayers, suitable for this task even under different experimental conditions. The sensitivity and accuracy of the approach through 2D' mode is on the same level as through the G mode, however, the clear advantage of the 2D' arises when doping effects are present in the sample.

INTRODUCTION

The ubiquity of strain in graphene [1] necessitates a continuous development of methods for its monitoring and correct assessment. On one hand, strain can be used to intentionally manipulate the electronic structure of graphene (the so-called strain engineering) to further boost the range of possible applications of this unique material,[2-4] on the other hand, strain fluctuations are a known cause for disorder in graphene devices.[5-7] For both reasons, the ability to control and quantify the strain in graphene and graphene-based devices is utterly essential.

Raman spectroscopy is undoubtedly the method of choice when it comes to a rapid, non-destructive, and versatile characterization of graphene samples, capable of delivering not only basic information about the number of layers or quality of the specimens, but also about the levels of charge-transfer doping or strain.[8-9] However, in spite of a great progress of Raman methodology and analysis of the acquired data, there are still ambiguities regarding the quantification of doping and strain, especially when these effects mingle. Several attempts to disentangle doping and biaxial strain from the Raman spectra have been reported, using the correlation of the G and 2D band frequencies ($\omega_G$ and $\omega_{2D}$, respectively).[10-13] While such an approach is quite simple and yet powerful in certain very specific cases, a great caution must be taken when interpreting beyond ideal conditions (a purely biaxial strain in graphene homogeneously adhered to an impurity-free substrate). In any case, at least to verify the assumptions about the state of the system, other parameters of the G and 2D bands have to be correlated, such as their widths ($\Gamma_G$ and $\Gamma_{2D}$, defined as full-width at half-maximum throughout the article), or their intensity ratio ($I_{2D}/I_G$, taken from areas). However, if uniaxial strain comes into the play, the usability of the widths and intensities of both G and 2D bands as well as the Raman shift of the 2D band becomes at least questionable, if not erroneous.

The G band corresponds to a first order Raman scattering process with a phonon of almost zero momentum. It is associated with the in-plane, doubly degenerate phonon from the transverse optical (TO) and longitudinal optical (LO) branches with $E_{2g}$ symmetry at the Brillouin zone center ($\Gamma$ point).[14] Both frequency and width of the G mode are strongly influenced by doping [15-16] and stress.[1, 17] The increase of $\omega_G$ in both electron- and hole-doped graphene is caused by a non-adiabatic removal of the Kohn anomaly at $\Gamma$ point,[18] and the simultaneous decrease of $\Gamma_G$ is caused by the Pauli blocking of phonon decay into electron-hole pairs. In general, mechanical stress causes G band downshift under tension and upshift under compression with a rate of 57 cm$^{-1}$/% for biaxial strain.[1] Under uniaxial strain the G bands splits into two components, G$^-$ and G$^+$, with the shift rates of ~ -31 and -10 cm$^{-1}$/%, respectively, for graphene on polymer substrates,[19-20] or ~ -37 and -19 cm$^{-1}$/%, resp., for suspended graphene.[21]

The 2D mode originates from a second-order triple resonant process between non-equivalent K points in the Brillouin zone of graphene, involving two zone-boundary, TO-derived phonons with opposite momenta $q$ and $-q$.[22-23] Several theoretical and experimental works [22, 24-28] address various aspects of the origin and properties of this intense and dispersive line, yet there are still several debatable issues. One of the crucial topics arises from the non-trivial shape of the 2D band in suspended graphene,[29] which is even more pronounced in strained samples.[27, 30-32] In a simplified one-dimensional portrayal of the 2D mode origin there are two dominant directions of the contributing phonon wavevectors – along K-$\Gamma$ (so called inner) or K-M (outer) symmetry lines in a k-space. Recent studies point to a greater contribution of the inner processes.[22, 30-31, 33] Nevertheless, recently, a full two-dimensional description of the electronic bands, phonon dispersion and matrix elements [22] was shown, in which the notation of inner or outer phonons is of a weaker relevance.[27] The 2D mode is dispersive, and its frequency changes with the excitation energy ($E_L$) with the slope $\partial\omega_{2D}/\partial E_L \sim 100$ cm$^{-1}$eV$^{-1}$.[9] The 2D band is also sensitive to doping and mechanical stress but these effects manifest themselves differently from those of the G band.[1] Strain causes the 2D band shift in the same directions as the G band, with the shift rates for biaxial strain larger by a factor of ~ 2.2-2.5.[34-35] Broadening and splitting of the 2D band under uniaxial strain is very different to that of the G band,[30] but no such effects are observed for biaxial deformation.[35] Doping causes increase of $\omega_{2D}$ for hole doping with a $\partial\omega_{2D}/\partial\omega_G \sim 0.5$-$0.7$,[10, 15, 36] whereas electron doping causes only a negligible $\omega_{2D}$ change for $|n| \leq 2 \times 10^{13}$ cm$^{-2}$ followed by a non-linear $\omega_{2D}$ decrease for higher n-doping levels.[15, 36] $\Gamma_{2D}$ increases and $I_{2D}$ decreases upon both p- and n-doping mainly due to electron-electron interactions, and also electron-phonon coupling strength.[22, 37]

In essence, the G band provides an acceptable measure of the uniaxial strain through the splitting into G$^+$ and G$^-$ components (at higher strain levels), but their Raman shifts may be swayed by doping, whereas the 2D band can serve well to determine smaller strain levels as it is not very influenced by doping. On the other hand, the usability of the 2D band at higher uniaxial strain levels is low due to the very complicated nature of its splitting. Therefore, it is highly desirable to find another feature in the Raman spectra, preferably more sensitive – in terms of frequency – to strain than to charge transfer, and in the same time, with a more defined response to uniaxial strain than the 2D mode. A promising candidate was proposed recently by Narula and Reich [38] – the 2D' mode. This feature arises from two LO-derived phonons from the vicinity of the Γ point, at ~ 3240 cm$^{-1}$, i.e. twice the frequency of the D' mode. The 2D' mode is yet less explored due to its much lower intensity compared with its renowned counterpart, the 2D mode.

Few studies of graphene under biaxial strain considered the 2D',[35, 39] however, none under uniaxial strain. Based on the insensitivity of the D' frequency to charge transfer,[40] one can assume only small effect of doping on the 2D' mode, either. Hence, based on the calculated behavior of the 2D',[38] which shows a robust and distinct splitting in a linear dependence on the strain level (for small strains), monitoring of this particular feature could suitably complement the strain assessment in graphene by Raman spectroscopy. However, the practical application of the 2D' for such a task has yet to be verified.

Here, we analyze the behavior of the Raman G, 2D and 2D' modes in uniaxially strained graphene on different substrates, with different adhesion levels, using lasers in a broad range of visible-light wavelengths for the excitation, to confirm the predicted theoretical behavior of the 2D' mode and thus its applicability to quantify strain in the system.

EXPERIMENTAL METHODS

The graphene samples were prepared by mechanical cleavage from Kish graphite and deposited on a clean and flexible polymethylmethacrylate (PMMA) substrate, previously spin coated with different polymers, SU8 photoresist (SU8 2000.5, MicroChem), PMMA (3% in anisole, MicroChem) and polycarbonate (PC) (3% in cyclopentanone, Sigma Aldrich). Single layer graphene flakes larger than 10 μm long (along the strain direction) were selected using an optical microscope. The samples were soft-cured at 80°C for 3 hours, and exposed to UV radiation (366 nm, 30 seconds) in the case of SU8. After curing the base polymer, some of the SU8 samples were covered with PMMA (1% in anisole), in order to improve the strain transfer efficiency from the flexible substrate to the graphene flake. Further in text, the uncovered samples are denoted as simply supported and the covered ones as fully supported. Tensile strain in the graphene was imposed through cantilever beam bending experiments, where the nominal measured strain on the flake ($\varepsilon_m$) was obtained from the expression:

$$\varepsilon_m = \frac{3t\delta}{2L^2}(1-\frac{x}{L}) \qquad (1)$$

where $t$ is the thickness of the substrate (3 mm), $\delta$ is the deflection of the bar, $L$ is the distance from the clamp to the deflection point (60 mm) and $x$ is the distance from the clamp to the flake position.

Raman spectra were recorded by LabRAM HR (Horiba Jobin-Yvon) and WITec Raman spectrometers, with a 600 grooves mm$^{-1}$ grating and a 100x objective. Lasers with various excitation energies were used: 2.71 eV (457.9 nm), 2.54 eV (488.0 nm), 2.41 eV (514.5 nm), 2.33 eV (532.0 nm), 2.18 eV (568.2 nm), and 1.96 eV (632.8 nm), while keeping the laser power below 1 mW to avoid heating of the polymer. The nominal strain ($\varepsilon_m$) was increased in 0.1 % steps and, additionally, for $\varepsilon_m$ above 0.5% an analyzer (in vertical and horizontal position) was placed before the grating to select the polarization direction of the scattered light in some experiments. One to several bending cycles were performed for each sample. Point-to-point spectral resolution varies for the two spectrometers and the excitation wavelength used, from ~ 1 to ~ 3 cm$^{-1}$.

RESULTS AND DISCUSSION

Figure 1 shows an example of the evolution of Raman spectra of a monolayer graphene flake simply supported (i.e., not covered by an additional polymer layer) on a PC layer spincoated onto a PMMA bar under uniaxial tension. As can be seen, all the graphene's bands redshift, as expected,[1, 20] and all of them split into at least two components. While the analysis of the splitting of the G band into $G^-$ and $G^+$ is quite straightforward and can be used to determine the locally imposed strain level (see below), the splitting of the 2D band into two components is only apparent [27, 30] and thus it might be questionable to utilize it for the quantification of higher strain levels.

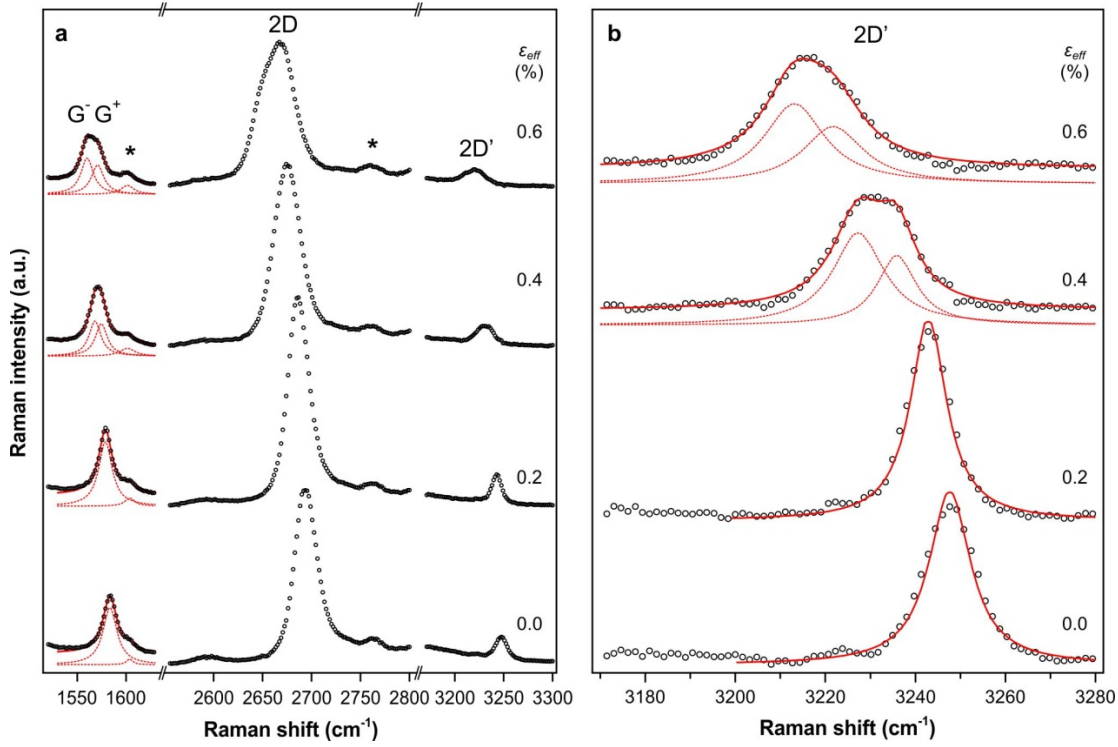

**Figure 1.** Evolution of the Raman spectra of simply supported monolayer graphene under tension, using a PMMA bar with a PC thin layer as the substrate. In (a) the original intensities of the spectra in all windows (G, 2D, and 2D') for a particular spectrum are retained, in (b) the 2D' spectra from (a) are enlarged. All spectra are offset for clarity. Points represent experimental data, red dashed lines individual Lorentzian lineshapes and red full lines their sum. Efficient strain ($\varepsilon_{eff}$), see below, is indicated. Laser excitation energy is 2.54 eV. Polarization direction is along the strain axis, no analyzer was employed. Asterisk denotes bands of the polymer.

Hence, at this moment, we will focus on the strain quantification using the shift rates of the G$^-$ and G$^+$ components. For the broad group of samples measured in the presented experiments, the shift rates $\partial\omega_{G^-}/\partial\varepsilon_m$ and $\partial\omega_{G^+}/\partial\varepsilon_m$ were found to fluctuate in the range from ~ -10 to -30 and from ~ -4 to -10 cm$^{-1}$/%, respectively, with their ratio, $\dfrac{\partial\omega_{G^-}/\partial\varepsilon_m}{\partial\omega_{G^+}/\partial\varepsilon_m}$, ranging between 2 and 2.5 for the simply supported samples, and up to ~3 for the fully supported samples. The range of shift rates with the value of their ratio of 3 can be expected considering recent works describing the varying stress transfer efficiency from the polymer matrix into the inclusion (graphene flake in our case) in the fully supported samples.[41-43] On the other hand, the lower values of $\dfrac{\partial\omega_{G^-}/\partial\varepsilon_m}{\partial\omega_{G^+}/\partial\varepsilon_m}$ for simply supported flakes indicate the presence of an additional factor influencing the stress transfer. The shifts of the G sub-bands can be expressed as:[20]

$$\Delta\omega_{G^\pm} = -\gamma_G \omega_G^0 (\varepsilon_{ll}+\varepsilon_{tt}) \pm \frac{1}{2}\beta_G \omega_G^0 (\varepsilon_{ll}-\varepsilon_{tt}) = -\gamma_G \omega_G^0 (1-\nu)\varepsilon \pm \frac{1}{2}\beta_G \omega_G^0 (1+\nu)\varepsilon \qquad (2)$$

where $\varepsilon_{ll}$ and $\varepsilon_{tt}$ are strain components parallel (longitudinal) and transverse to the strain axis, $\gamma_G$ and $\beta_G$ are the G mode's Grüneisen parameter and shear deformation, respectively, and ν is the Poisson ratio. For fully supported samples, the shift rates are governed by the Poisson ratio of the polymer, taken as 0.33, which gives ~ -31 and -10 cm$^{-1}$/% for $\partial\omega_{G^-}/\partial\varepsilon$ and $\partial\omega_{G^+}/\partial\varepsilon$, resp., for a full stress transfer.[1, 20] From that, $\gamma_G$ = 1.99 and $\beta_G$ = 0.99. However, for suspended graphene, the Poisson ratio of graphene has to be considered. Using the above mentioned $\gamma_G$ and $\beta_G$, one gets $\partial\omega_{G^-}/\partial\varepsilon$ ~ -36.4 cm$^{-1}$/% and $\partial\omega_{G^+}/\partial\varepsilon$ ~ -18.6 cm-1/% for free graphene.[20] These values have been experimentally confirmed recently.[21] The ratio between the shift rates drops below 2 in this case. Given the range of ratios measured in our experiments on simply supported samples (~ 2-2.5) combined with the lower absolute shift rates, it is thus reasonable to assume that the smaller stress transfer in the uncovered samples is also accompanied by a change in the governing Poisson ratio from the polymer's to graphene's, on a sub-microscopic scale, originated in non-adhering portions of the flakes. In that case, one cannot obtain the stress transfer efficiency ($\eta$) by simply comparing the measured shift rates to the referenced values of -31 and -10 cm$^{-1}$/%,[20] but the maximum shift rates have to be recalculated using the particular $v_m$ obtained experimentally, using the ratio between the shift rates for $\varepsilon_m$:

$$v_m = \frac{\frac{\partial \omega_{G^-}/\partial \varepsilon_m}{\partial \omega_{G^+}/\partial \varepsilon_m}(\gamma_G - \frac{1}{2}\beta_G) - (\gamma_G + \frac{1}{2}\beta_G)}{\frac{\partial \omega_{G^-}/\partial \varepsilon_m}{\partial \omega_{G^+}/\partial \varepsilon_m}(\gamma_G + \frac{1}{2}\beta_G) - (\gamma_G - \frac{1}{2}\beta_G)} \quad (3)$$

For each particular $v_m$ the maximum theoretical shift rate can be calculated from equation (2) and the stress transfer efficiency $\eta$ calculated as the ratio between the measured and theoretical shift rates. Finally, for each measured nominal strain, at which a Raman spectrum was acquired, one can obtain the efficient strain $\varepsilon_{eff} = \eta \varepsilon_m$. In Figure 1, the values of $\varepsilon_{eff}$ are already given.

The 2D' band redshifts too and can be fitted by two Lorentzian components, as predicted by Narula and Reich.[38] Figure 1b shows the enlarged 2D' band region from Figure 1a, including the fitting by Lorentzian lineshapes. The splitting is visible (and can be fitted with a reasonable goodness) from ~0.3%; the ratio between the two components remains the same for a particular sample with a given polarization setup, and their efficient shift rates are ~ -60 to -65 and ~ -40 to -45 cm$^{-1}$/%, for the faster and slower moving component, respectively, again in line with the calculations from.[38] In Figure 2, $\partial \omega_{G^-}/\partial \varepsilon$ and $\partial \omega_{G^+}/\partial \varepsilon$ correspond to $v = 0.13$, the shift rates with the 95% confidence interval being -37.1 ± 1.0 and -18.6 ± 0.7 cm$^{-1}$/% for G$^-$ and G$^+$, respectively. For 2D', the shift rates in Figure 2 are -63.2 ± 1.9 and -42.7 ± 2.6 cm$^{-1}$/%. As can be seen, the larger spread of the 2D' values corresponds to its larger frequency; but given its approximately twice larger strain sensitivity, the precision of thus derived strain would be equal to the strain derived from the G sub-band's shifts. The larger spread of the fitted data for 2D' is mostly due to its lower intensity.

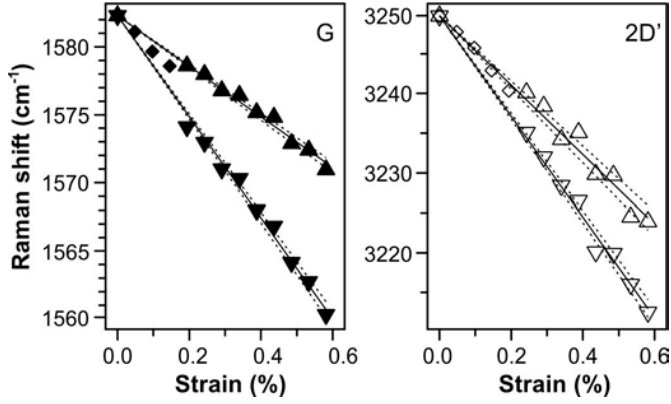

**Figure 2.** Evolution of the individual components of the G (left) and 2D' (right) bands from Figure 1 as fitted by one (diamonds) or two (triangles) Lorentzian lineshapes. Full lines are the least squares line fits to the experimental data, while the dashed lines define the confidence intervals (95%) for the respective line fits. Laser excitation energy is 2.54 eV.

The frequency shift of the 2D' mode with uniaxial strain can be expressed in analogy with the $G^+$ band, as they both are associated with the LO phonon at Γ, *cf.* equation (1):[38]

$$\omega_{2D'} = \omega_{2D'}^0 - \gamma_{2D}\omega_{2D'}^0(1-\nu)\varepsilon + \frac{1}{2}\beta_{2D}\omega_{2D'}^0(1+\nu)\varepsilon \qquad (4)$$

The response of the 2D' band is highly sensitive to the orientation of the crystal lattice with respect to the strain axis ($\varphi$) and to the directions of the incident and scattered light polarization ($\theta_{in}$ and $\theta_{out}$, respectively), owing to the fluctuations of $\beta_{2D'}$ for various $\varphi$: $\theta_{in}$: $\theta_{out}$ combinations.[38] The largest difference between the calculated frequencies of 2D' band maxima is achieved when $\theta_{in} = \theta_{out}$, i.e., in parallel polarization configuration, and $\varphi$ changes from 0 to 90°, i.e. from the zigzag to armchair edge oriented along the strain direction. However, as can be seen in the example in Figure 1 and 2, not employing the analyzer leads to a 2D' shape, which envelops the response of phonons scattered at all polarization angles. Consequently, fitting this convoluted shape with two Lorentzian components provides the frequencies matching the 2D' band at the minimum and maximum shift under the extreme polarization conditions. In the example in Figure 2 (right), the difference between the two components ($\partial \Delta \omega_{2D'} / \partial \varepsilon$) amounts to 20.5 cm$^{-1}$/%, very close to the value of 21 cm$^{-1}$/% calculated by Narula and Reich.[38] Two drawbacks are associated with such an approach: (i) as mentioned above, the two components can be fitted with a reasonable level of goodness only from a certain strain level, and (ii) under certain polarization/lattice orientation/strain direction conditions, the two components may not necessarily be as far apart as they would be when the extreme conditions are met.

Indeed, it is possible to conduct the experiment in the way where $\theta_{in} = \theta_{out}$ and they are rotated around 90° along the strain axis direction. An example of such a pair of spectra is shown in Figure 3 (simply supported flake on SU8 photoresist thin film on a PMMA beam). The difference between the frequencies of the 2D' band is ~ 9 cm$^{-1}$ at $\varepsilon_{eff}$ = 0.4%, which gives $\partial \Delta \omega_{2D'} / \partial \varepsilon$ of ~ 22.5 cm$^{-1}$/%, again very close to the predicted difference.

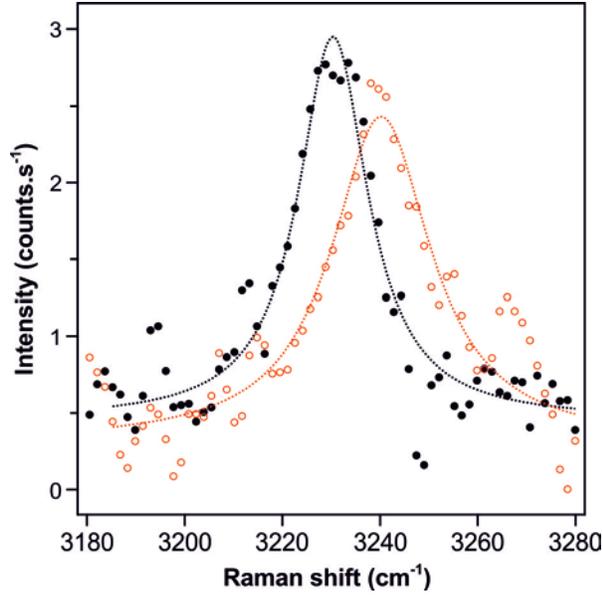

**Figure 3.** 2D' Raman band of uniaxially strained simply supported monolayer graphene on SU8 photoresist on a PMMA beam at $\varepsilon_{eff}$ = 0.4% at parallel (red) and perpendicular (black) direction of the polarizer/analyzer with respect to the strain axis. Laser excitation energy is 2.54 eV.

Unfortunately, the presence of the λ/2 waveplate and the analyzer, which take up a considerable amount of transmitted light, results in a much lower signal-to-noise ratio of the spectra. This, combined with the low sensitivity in the 2D' region of standard CCD cameras optimized for visible wavelengths, and the necessity to keep the laser power and/or exposure time at a level that does not damage the heat-sensitive polymer, makes the precision of such measurements drop to a degree similar to the above described procedure without the analyzer (and the λ/2 waveplate).

In order to test the robustness of the strain quantification through the difference both between the 2D' bands under the different polarization conditions, and between the two components, when all scattered polarizations are allowed (as the precisions of the two methods are alike), we measured a broad set of samples on different substrates (SU8, PMMA, PC), both covered or uncovered, in some cases in several subsequent stretching cycles, with lasers of different excitation wavelengths, on two different Raman spectrometers (see Experimental for the whole range of conditions). For every step in each experimental run the efficient strain values were calculated using the G band splitting described above and the $\Delta_{2D'}$ values were obtained. The results are plotted in Figure 4. The least squares line fit to the experimental data gives $\partial \Delta \omega_{2D'} / \partial \varepsilon = 21.7 \pm 0.6$ cm$^{-1}$/%, hence very close to the calculated value of 21 cm$^{-1}$/%.[38]

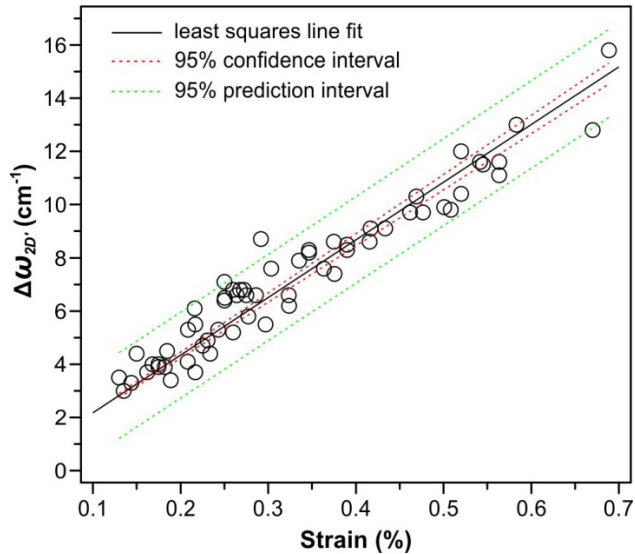

**Figure 4.** $\Delta \omega_{2D'}$ as a function of effective strain in uniaxially strained monolayer graphene flakes in experimental runs under varied conditions (different substrates, simply/fully supported, different laser excitations, different spectrometers). Circles are experimental points, black full line their least squares line fit with [0,0] intercept, red and green dashed lines marking the 95% confidence and prediction intervals, respectively.

CONCLUSIONS

We have measured an assortment of tensile-stress experiments on monolayer graphene flakes under various conditions, consisting of different substrates, both under- and overlying the specimens, different laser excitation energies, differently repeated cycles, and two different Raman spectrometers. Through the introduced analysis of the splitting of the G band into the G$^-$ and G$^+$, their shift rates with strain and the ratio of the shift rates, we obtained values of efficient strain at each measurement point. In the same time, we monitored the evolution of the 2D'band under different polarization conditions – either by setting the polarizer/analyzer into a selected combination and changing it from 0 to 90°, or by detecting the scattered light at all angles. In the former case, the difference between two variously shifted narrow 2D' bands at the opposite angles was measured. In the latter case, the broader and visibly asymmetric 2D'band was fitted with two Lorentzian lineshapes and the difference between their positions was obtained. In both cases, similar values of $\Delta_{2D'}$ were retrieved. From all the experimental data combined, a robust value $\partial\Delta\omega_{2D'}/\partial\varepsilon$ of 21.7 ± 0.6 cm$^{-1}$/% was extracted, very close to the predicted one of ~ 21 cm$^{-1}$/%.[38] The so obtained value of $\partial\Delta\omega_{2D'}/\partial\varepsilon$ provides a similar sensitivity as the G$^-$, G$^+$ components. Although the stress quantification using the 2D' band seems to be at the same level as using the G band, a clear advantage of using the 2D' band arises in doped graphene. If, for example, charge inhomogeneities from the substrate (charge puddles) are present in the system and uniaxial strain is imposed to control the electronic structure of the sample, the variation of both strain and doping would render the strain quantification through the G band hardly possible. Doping (dedoping) would induce the same shift to both G$^-$, G$^+$ components to higher (lower) frequencies and without knowing the stress transfer efficiency, the analysis through $\partial\omega_{G^-}/\partial\varepsilon_m$, $\partial\omega_{G^+}/\partial\varepsilon_m$ and their ratio would introduce errors into the strain quantification. While we cannot fully exclude that doping changes took place during some of our experiments, the varied conditions we used, especially regarding the different substrates (both bottom and top), should have provided enough statistically independent results to secure a reliable value of $\partial\Delta\omega_{2D'}/\partial\varepsilon$.

On top of that, since the value the 2D' splitting is used for the strain level quantification, and not the shift rates of the components, the doping effects should not influence this value even if present. The close proximity of our value, incl. a small 95% confidence interval, to the one predicted in [38] points to the viability of both the presented experimental and previous theoretical [38] results.

AUTHOR INFORMATION

**Corresponding Author**

* Otakar Frank, e-mail: otakar.frank@jh-inst.cas.cz, tel: +420 266053446, address: J.Heyrovsky of Institute of Physical Chemistry of the AS CR, v.v.i., Dolejskova 2155/3, 182 23 Prague 8, Czech Republic.**Author Contributions**

The manuscript was written through contributions of all authors. All authors have given approval to the final version of the manuscript.

ACKNOWLEDGMENT

This work was funded by Czech Science Foundation project No. 14-15357S and European Union FP7 Programme (No. 604391 Graphene Flagship). M.K. and E.d.C. acknowledge the support from ERC-CZ project No. LL1301.15

TOC Image

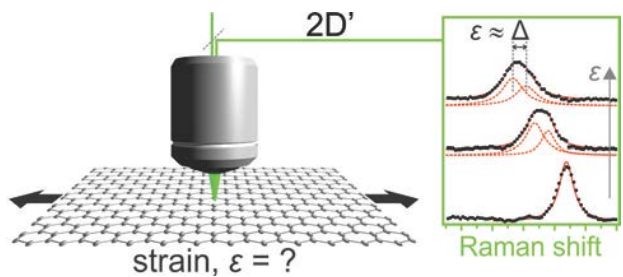